\documentclass[twocolumn,showpacs,preprintnumbers,amsmath,amssymb]{revtex4}

\usepackage{graphicx}
\usepackage{dcolumn}
\usepackage{bm}
\usepackage{amsmath}
\usepackage{amssymb}
\usepackage{latexsym}
\usepackage{epsfig}
\usepackage{amsbsy}
\usepackage{array}
\usepackage{amssymb}
\usepackage{setspace}
\usepackage{bm}

\begin{document}

\preprint{}

\title{ The Stability and Charge Carriers in Bilayer Silicene  }

\author{Wang Rui\footnote{Email: rcwang@cqu.edu.cn.}, Wang Shaofeng, and Wu Xiaozhi}
\affiliation{Institute for Structure and Function and
Department of physics, Chongqing University, Chongqing 400044, People's Republic of China.}

\date{\today}

\begin{abstract}

The structure optimization, phonon, and \emph{ab initio} finite temperature molecular dynamics calculations have been performed to predict that bilayer silicene has stable structure with AB stacking geometry and is more favorable energetically to synthesize than monolayer silicene, a two-dimensional honeycomb lattice with buckled geometry. Marvellously, its electronic bands show that the charge carriers behave like relativistic Dirac fermions with linear energy dispersions near the $K$ points. An insightful analysis has been presented to understand the low-energy electronic excitations based on tight-binding approximation, and we suggest that the component of $sp^{3}$ hybridization in the buckled geometry blocks the interlayer hopping, so the linear dispersion can be preserved.
\end{abstract}

\pacs{73.22.-f, 61.48.De, 63.22.-m}

\keywords{Suggested keywords}

\maketitle

Monolayer graphene, a two-dimensional (2D) honeycomb network of carbon atoms, has attracted great attention because of the unusual electronic properties like the massless Dirac fermions \cite{Geim,Neto}. Not only have the unique properties of monolayer graphene been payed close attention, but also the geometric, electronic \cite{Latil,McCann}, and vibrational properties \cite{Yan} of bilayer or few-layer graphene have been investigated intensely. Inspired by the fruitful results of graphene, the honeycomb structure of Si (silicene) has been predicted to be stable theoretically \cite{Cahangirov}.  Unlike the graphene with planar geometry, silicene has the low-buckled geometry, in which there is a buckling distance with the two sublattices of the honeycomb lattice being displaced vertically with respect to the other \cite{Cahangirov,Jose}.

Recently, the synthesis of silicene sheets on a Ag (111) substrate had been realized in several groups exhilaratingly \cite{Vogt, Feng, Chen}.
Most importantly, the charge carriers in monolayer monolayer silicene can behave like a massless Dirac fermions because of their valence ($\pi$) bands and conduction ($\pi^*$) bands crossing linearly at the Fermi level.  Due to its compatibility with Si$-$based electronics and potential applications in future nanoelectronic devices, there has been a lot of theoretical interest in the low-energy electronic and related properties \cite{Drummond, Houssa1,Houssa2,Cahangirov2010,Ding,Bechstedt,Kang}. Similar to grapehen, the bilayer as well as multilayer silicene  would also like to exhibit unusual physical properties \cite{Bai, Padova,Ezawa}.

In this letter, we demonstrate a stable structure of bilayer silicene based on density functional theory (DFT), while the state-of-the-art structure optimization, phonon, and \emph{ab initio} finite temperature molecular dynamics calculations have been carried out. The results show that the bilayer silicene with AB stacking can maintain the stable 2D lattice and is predicted to be synthesized more easily than monolayer silicene experimentally. It is very interesting that the electronic band structures of bilayer silicene with AB stacking somehow exhibit the linear energy dispersion relations near the $K$ and $K'$ points in the hexagonal Brilouin zone (BZ). In order to understand the low-energy electronic excitations of the massless Dirac fermions in bilayer silicene, a insightful analyzing has been performed within the framework of tight-binding approximation. The characteristics of low-energy excitation, which offer exciting opportunities for the occurrence of new phenomena and the development of high performance electronic devices, have also been discussed.

The DFT calculations calculations are carried out with the local density approximation (LDA) \cite{Ceperley} and the projector-augmented wave (PAW) potentials \cite{Blochl,Kresse1}, implemented as the Vienna \emph{ab initio} simulation package (VASP) \cite{Kresse2}. A plane-wave basis set with energy cutoff of 500 eV is used. The BZ is sampled by ($30\times30\times1$) special $\mathbf{k}$ points for the total-energy and electronic-structure calculations. We assume a slab geometry with a vacuum spacing of 15 {\AA} in order to simulate the 2D crystal. The atomic structure optimization has been carried out by relaxing the forces on all the atoms using a conjugate gradient method. The convergence for energy and Hellmann-Feymann force between two steps are chosen as $10^{-7}$ eV and $10^{-4}$ eV/{\AA}. The phonon dispersions have been calculated  by using a supercell approach \cite{Parlinski}. The density-functional perturbation theory (DFPT) have been performed to obtain real-space force constants of supercells by VASP, and phonon frequencies have been calculated from the force constants in 7$\times$7 supercell as implemented in PHONOPY package \cite{Togo}.

Based on the symmetry of the honeycomb structures with low-buckled geometry, there are two structures [AA stacking (FIGs.\ref{fig1}a) and AB stacking (FIGs.\ref{fig1}b)] that can form the stable bilayer silicene potentially, as shown in FIG.\ref{fig1}. The geometry optimization has been performed through variation of the lattice constant and full relaxing the atomic coordinates. The results show that both AA stacking and AB stacking bilayer can arrive at the minima of energy locally, and the buckled honeycomb structures of per layer of silicene have also been preserved. Furthermore, the structure relaxation demonstrates that per layer of bilayer silicene has the equivalent value for the buckling parameters $t$, due to the symmetry of space inversion.

\makeatletter
\def\@captype{figure}
\centerline{\scalebox{0.5}[0.5]{\includegraphics{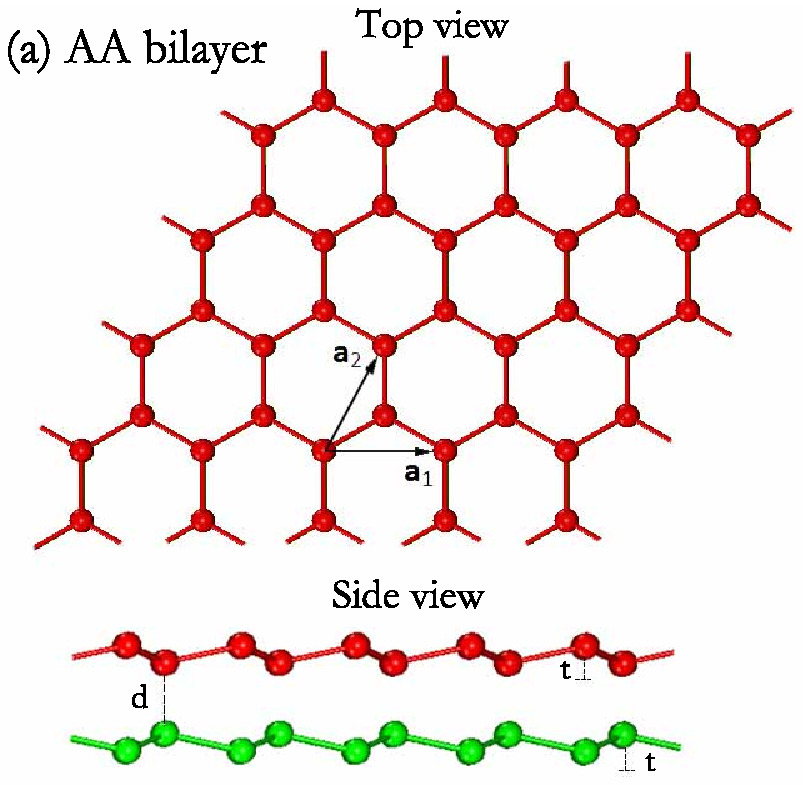}}}
\centerline{\scalebox{0.5}[0.5]{\includegraphics{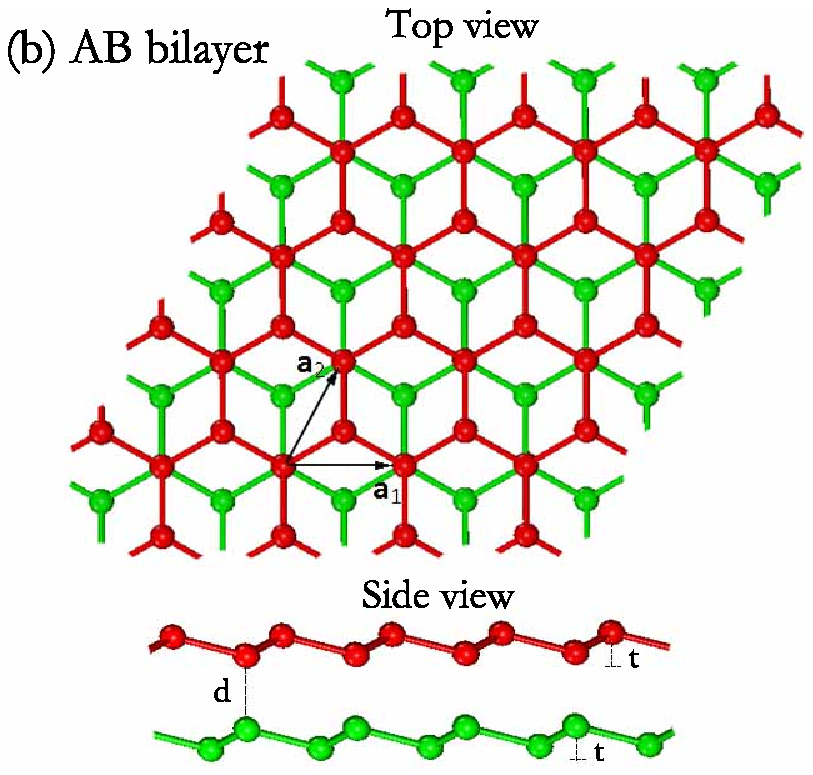}}}
\makeatother
\caption{\footnotesize{(Color online) Top view and Side view of bilayer silicene with (a) AA stacking and (b) AB stacking. The red pattern and green pattern represent the upper and lower layer of silicene, respectively. $\mathbf{a}_{1}$ and $\mathbf{a}_{2}$ denote the hexagonal lattice vectors, $d$ indicates the distance between two layers, and $t$ denotes the buckled distance.} }\label{fig1}
Table \ref{tab:structure} lists the structural parameters and cohesive energies for the bilayer silicene with AA stacking and AB stacking in comparison with the monolayer silicene. The lattice parameters and cohesive energies of monolayer silicene agree well with the previous calculations \cite{Cahangirov}. The buckling $t_{AA}=0.647$ {\AA} and $t_{AB}=0.659$ {\AA} of AA stacking and AB stacking bilayer silicene are slightly larger than that of monolayer silicene. In the 2D honeycomb silicene, the buckling geometry is formed by mixed $sp^2$ and $sp^3$ hybridization. Hence, the bonds of bilayer silicene contain more components of $sp^3$ hybridization than monolayer silicene. The bilayer siliene with AB stacking shows the lowest cohesive energies $E_{c}=5.32$ eV, so it is substantially more stable than AA stacking($E_{c}=5.25$ eV) and monolayer silicene ($E_{c}=5.13$ eV).

\begin{table}[h]
\caption{The optimized structural properties and cohesive energies of monolayer and bilayer silicene. The lattice constant $a=|\mathbf{a_{1}}|=|\mathbf{a_{2}}|$, buckling distance $t$, interlayer spacing $d$, and cohesive energy $E_{c}$ are calculated from the PAW-LDA. }
\begin{center}
\begin{tabular}{lllllllllll} 
\hline
\hline
 &  $a$(\AA) & $t$ (\AA) & $d$ (\AA) & $  E_{c}$ (eV)  \\
\hline
monolayer & 3.825, 3.83\cite{Cahangirov} & 0.438, 0.44\cite{Cahangirov} & \ \ - &\ \ 5.13, 5.16\cite{Cahangirov}  \\
AA layer  & 3.817 & 0.647 & 2.429 &\ \ 5.27  \\
AB layer  & 3.806 & 0.659 & 2.481 &\ \ 5.32  \\
\hline \hline
\end{tabular}
\end{center} \label{tab:structure}
\end{table}

The phonon dispersion curves for bilayer silicene with AA stacking and AB stacking are shown in FIG. \ref{fig2}. It can be seen that the bilayer silicene with AA stacking has the imaginary phonon frequencies that imply the crystal is unstable, while the AB bilayer silicene shows lattice stabilities because of the calculated phonon modes all being positive over the BZ. In fact, the imaginary frequencies of AA bilayer silicene result from the modes of relative vibration of interlayer, as ZO branches (so-called out-of-plane optical branch), while the ZO branches can maintain the dynamical equilibrium for the bilayer with AB stacking.  Furthermore, we have carried out \emph{ab initio} finite temperature molecular dynamics calculations to test the stability of 2D bilayer silicene with AB stacking with time steps of 2 fs. The $7\times7$ supercell containing 196 atoms is used to lift the constraint of unit cell. By rasing the temperature from $T=0$ to $700$ K in 50 steps and keeping the temperature of the system around 700 K for 2.5 ps, the 2D bilayer honeycomb geometry of AB stacking is not destroyed. This convinces us that,
as a metastable 2D crystal, the bilayer silicene with AB stacking can be stable.

\makeatletter
\def\@captype{figure}
\centerline{
\scalebox{0.24}[0.24]{\includegraphics{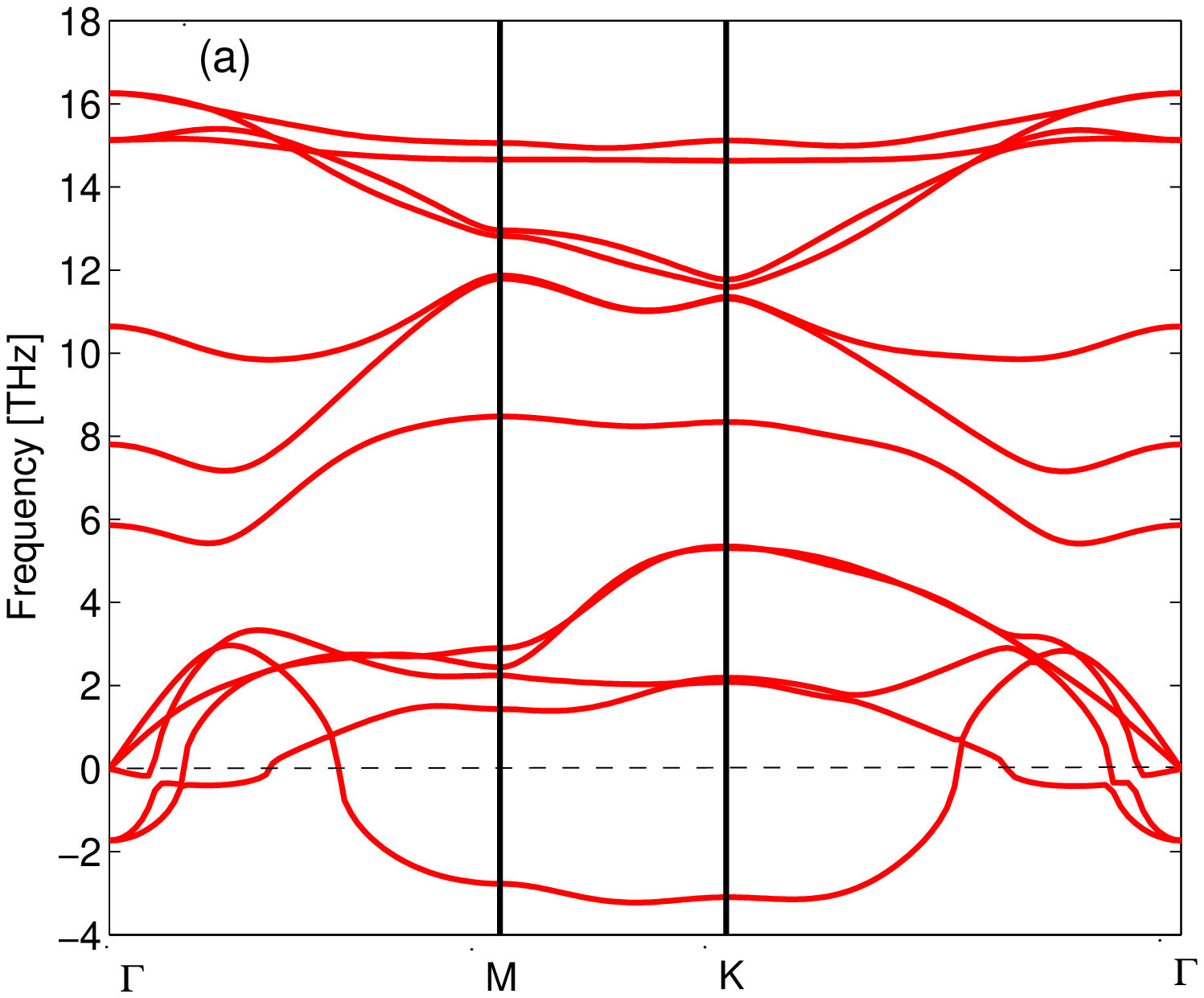}}
\scalebox{0.24}[0.24]{\includegraphics{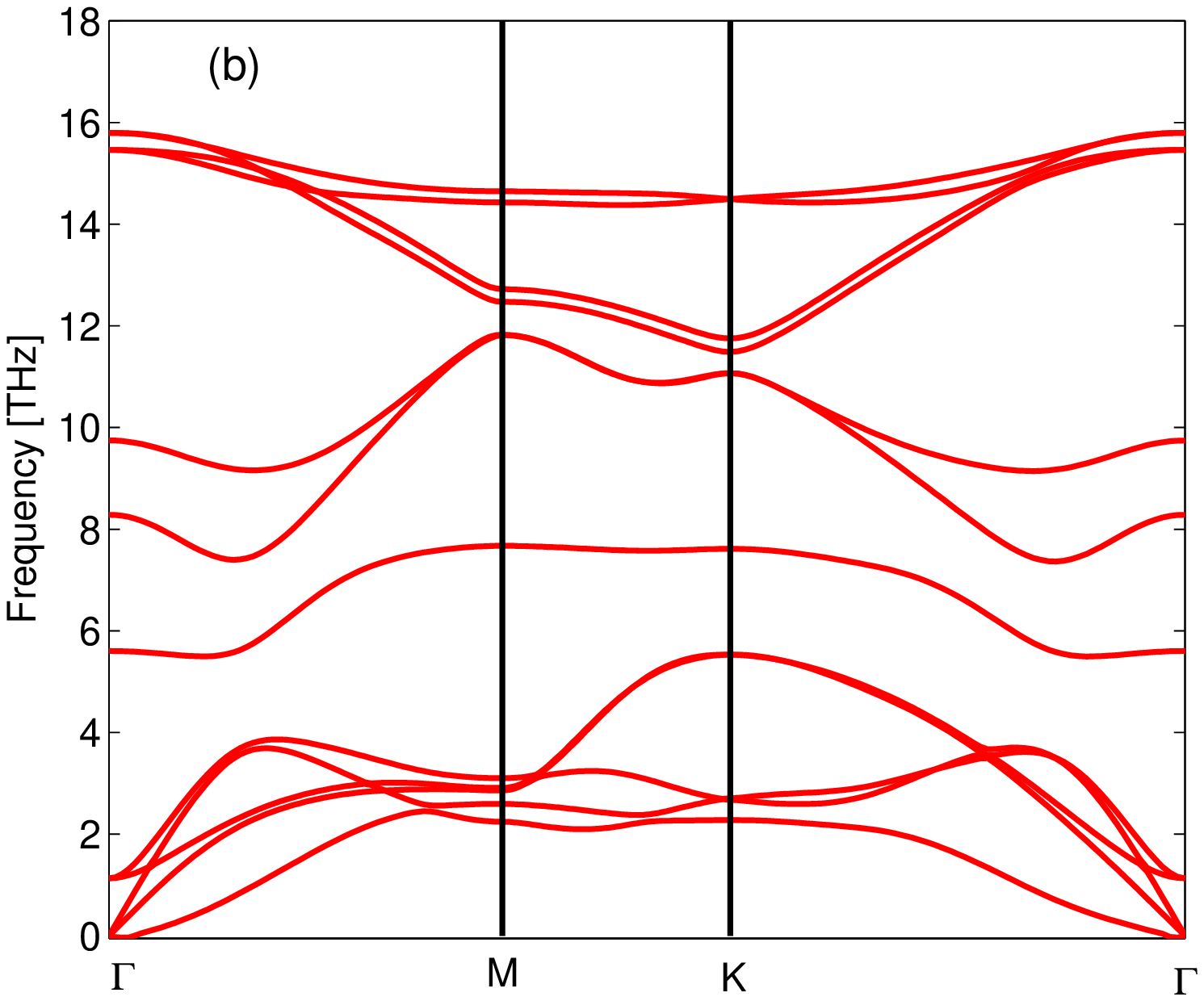}}}
\makeatother
\caption{\footnotesize{(Color online) Phonon-dispersion curves calculated by
supercell approach for 2D bilayer silicene (a)AA stacking and (b) AB stacking along the high-symmetry
line in the BZ.} }\label{fig2}

The calculated electronic band structures of bilayer silicene with AB stacking are shown in FIGs. \ref{fig3}b, and the band structures of monolayer silicene are also reproduced in FIGs. \ref{fig3}a for comparison. The bands of monolayer and AB bilayer silicene have the similar framework near the Fermi level. Similar to monolayer silicene,  it is remarkable to note that $\pi$ and $\pi^{*}$ bands of AB bilayer silicene cross linearly at $K$ and $K'$ points at $E_{F}$. The results mean that the charge carriers in AB bilayer silicene are the massless Dirac fermions. The Fermi velocity $v_{F}$ can be estimated by fitting the $\pi$ and $\pi^{*}$ bands close to the $\mathbf{K}$ (or $\mathbf{K}'$) vector as $\mathbf{k}=\mathbf{K}+\mathbf{q}$, with $|\mathbf q|\ll|\mathbf K|$, to the expression $E(\mathbf q)\simeq \hbar v_{F}|\mathbf q|$, and then the results are $v_{F}^{\mathrm{m}}\simeq 1.15 \times 10^{6}$ m/s for monolayer silicene and $v_{F}^{\mathrm{b}}\simeq 0.65\times 10^{6}$ m/s for AB bilayer silicene.

\makeatletter
\def\@captype{figure}
\centerline{
\scalebox{0.24}[0.24]{\includegraphics{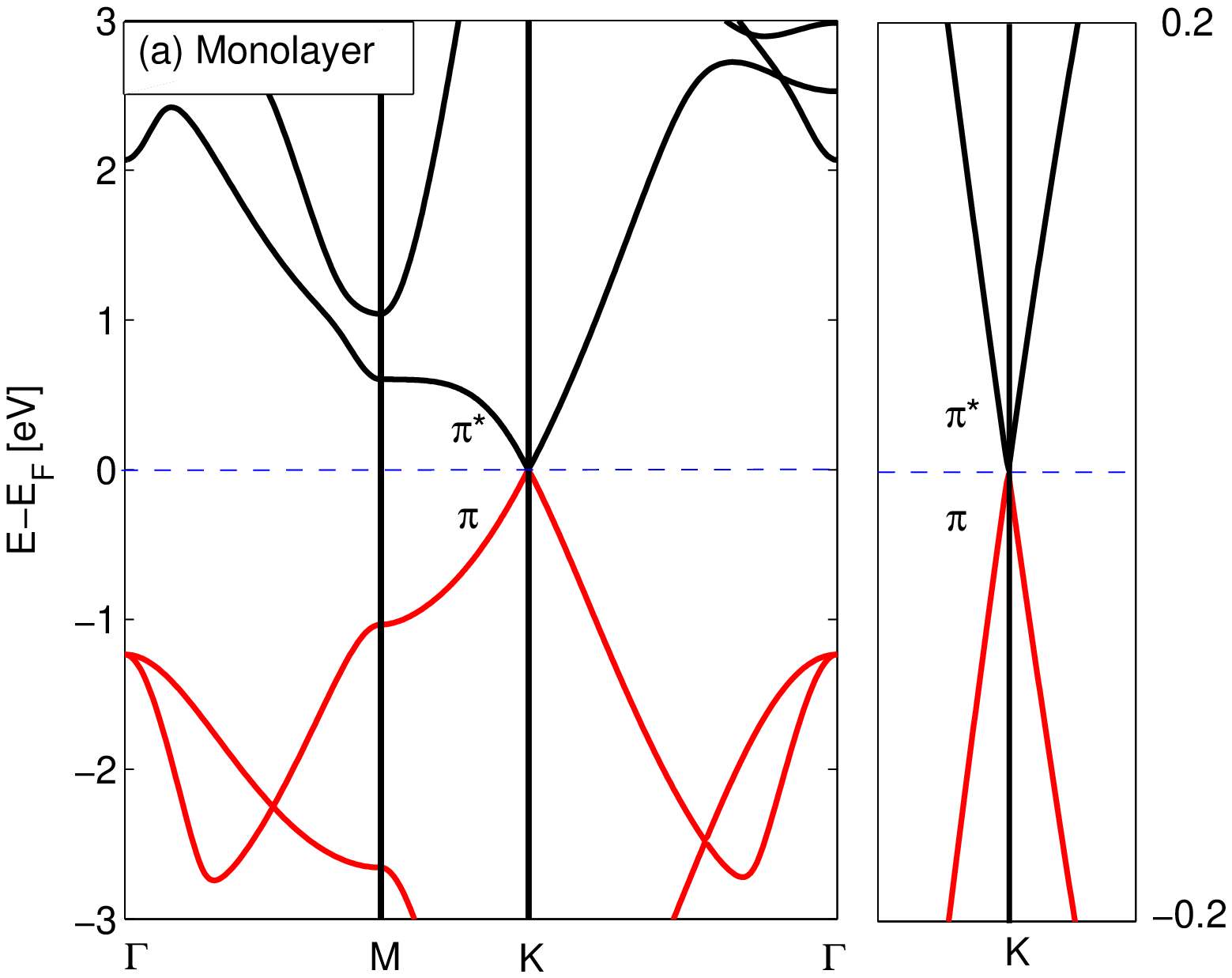}}
\scalebox{0.24}[0.254]{\includegraphics{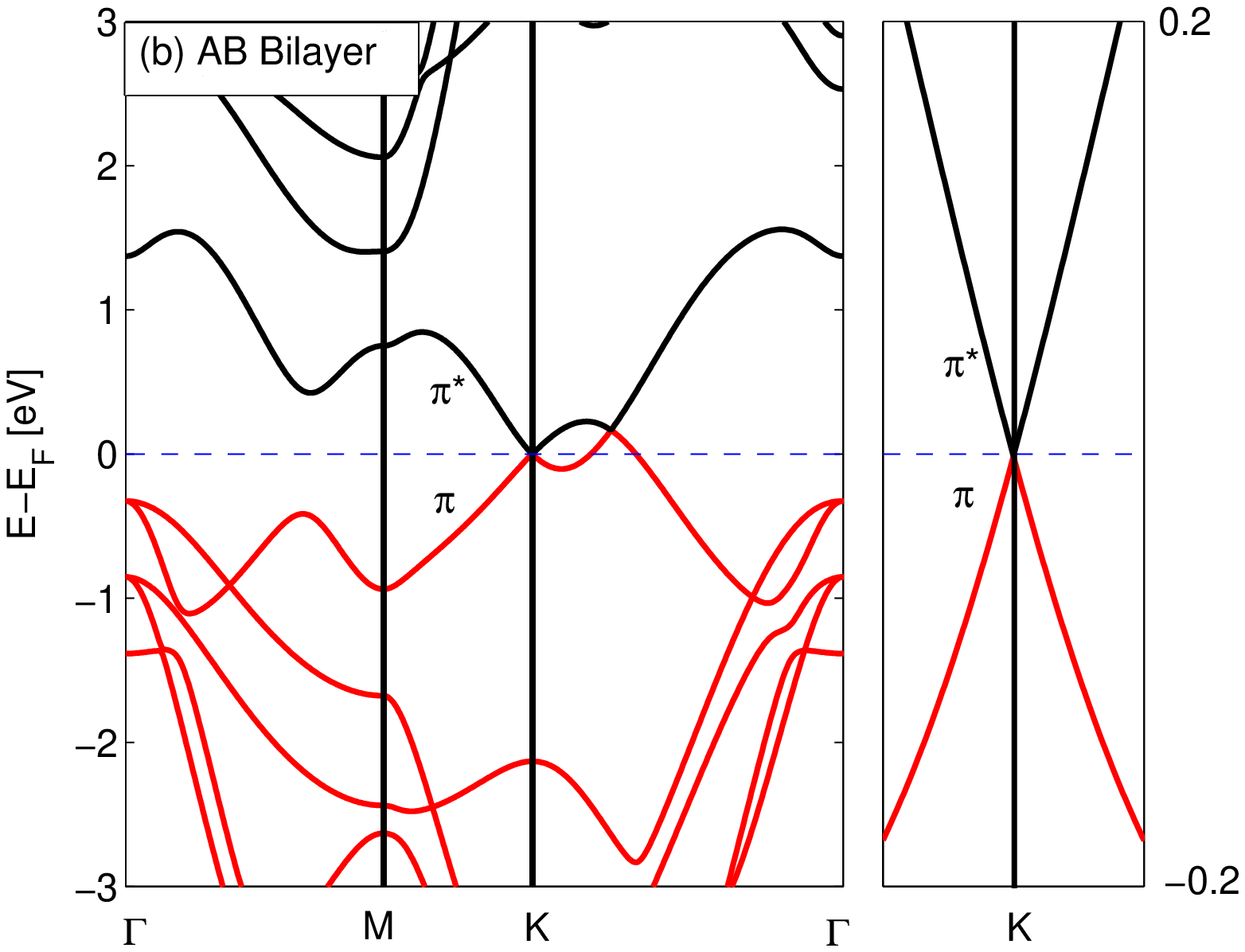}}}
\makeatother
\caption{\footnotesize{(Color online) The band structures of (a) monolayer silicene and (b) bilayer silicene with AB stacking show the remarkable linear dispersion of $\pi$ and $\pi^{*}$ bands at Fermi level. The dispersion-relations near the $K$ point are magnified at the right side.} }\label{fig3}

We construct an effective 2D Hamiltonian from the tight-binding (TB) approximation to understand the low-energy electronic excitations of charge carriers in AB bilayer silicene. Considering the electrons hopping to nearest neighbor atoms in-layer and interlayer (see FIGs.\ref{fig4}a), the TB Hamiltonian has the form
\begin{eqnarray}\label{hamiltonian-real}
H=&&-\gamma_{0}\sum_{\langle i,j\rangle,m,\sigma}\big(C_{a,m,i,\sigma}^{\dag}C_{b,m,j,\sigma}+\mathrm{H.c.}\big) \nonumber \\
&&-\gamma_{1}\sum_{{i}, \sigma}\big(C_{a,A,i,\sigma}^{\dag}C_{a,B,i,\sigma}+\mathrm{H.c.}\big) \nonumber \\
&&-\gamma_{2}\sum_{{i},\sigma}\big(C_{a,A,i,\sigma}^{\dag}C_{b,B,i,\sigma}
+C_{b,A,i,\sigma}^{\dag}C_{a,B,i,\sigma}+\mathrm{H.c.}\big), \nonumber \\
&&
\end{eqnarray}
where the operator $C_{\mu,m,i,\sigma}^{\dag}$ ($C_{\mu,m,i,\sigma}$) creates (annihilates) an electron with spin $\sigma(\sigma=\uparrow, \downarrow)$, on sublattice $\mu=a, b$, in layer $m=A, B$, at site $\mathbf{R}_{i}$. $\gamma_{0}$ is the nearest neighbor $\langle i,j\rangle$ in-layer hopping energy, $\gamma_{1}$ is the vertical interlayer hopping energy between atom $A_{a}$ and atom $B_{a}$, and $\gamma_{2}$ is the nearest neighbor interlayer hopping energy between atom $A_{a} (B_{a})$ and atom $B_{b}(A_{b})$. Conveniently, Hamiltonian (\ref{hamiltonian-real}) is transformed into momentum space by Fourier transformation as
\begin{eqnarray}
 C_{\mu,m,i}=\frac{1}{\sqrt{N_{\mu}}}\sum_{\mathbf{k}}\mu_{m}(\mathbf{k})e^{i\mathbf{k}\cdot \mathbf{R}_{i}}.
\end{eqnarray}
 In the long-wave approximation, by expanding the momentum close to $K$ point in the BZ, the Hamiltonian reads
\begin{equation}
H=-\sum_{\mathbf{k}}\Psi(\mathbf{k})^{\dag}\cdot \mathbf{h} \cdot \Psi(\mathbf{k}),
\end{equation}
where
\begin{equation}\label{h}
\mathbf{h }\simeq\left(
             \begin{array}{cccc}
               0 & v_{F}\Pi & \gamma_{1} & \xi \Pi^\dag \\
               v_{F}\Pi^\dag & 0 & \xi \Pi^\dag &0 \\
               \gamma_{1} & \xi \Pi  & 0 &  v_{F}\Pi^\dag \\
               \xi \Pi & 0 & v_{F}\Pi & 0 \\
             \end{array}
           \right),
\end{equation}
and $\Pi=q_x+iq_y$, $v_{F}=\frac{1}{\hbar}\frac{\sqrt{3}}{2}a\gamma_{0}$, and $\xi=\frac{1}{\hbar}\frac{\sqrt{3}}{2}a\gamma_{2}$, and
\begin{equation}
\Psi(\mathbf{k})^{\dag}=[a_{A}^{\dag}(\mathbf{k}), b_{A}^{\dag}(\mathbf{k}), a_{B}^{\dag}(\mathbf{k}),b_{B}^{\dag}(\mathbf{k})]
\end{equation}
is the four component spinor. Then, the energy-dispersions around the $K$ points can be determined from the eigenvalues of Eq. (\ref{h}) as
\begin{eqnarray}
&&E_{1}^{\pm}(\mathbf{q})\simeq \pm\gamma_{1}\pm \frac{2{v_{F}^{\pm}}^2  q_{0} }{ \sqrt{\gamma_{1}^2\pm  4{v_{F}^{\pm}}^2 q_{0}^2}}\Bigg|_{q_{0}=0}|\mathbf q| +\frac{{v_{F}^{\pm}}^2 }{ \gamma_{1}}|\mathbf q|^2, \nonumber \\
&&E_{2}^{\pm}(\mathbf{q})\simeq \mp \frac{2{v_{F}^{\pm}}^2q_{0} }{ \sqrt{\gamma_{1}^2+  4{v_{F}^{\pm}}^2  q_{0}^2}}\Bigg|_{q_{0}=0}|\mathbf q| \mp\frac{{v_{F}^{\pm}}^2 }{ \gamma_{1}}|\mathbf q|^2,
\end{eqnarray}
where $v_{F}^{\pm}=v_{F}\pm\xi$.

\makeatletter
\def\@captype{figure}
\centerline {\scalebox{0.18}[0.18]{\includegraphics{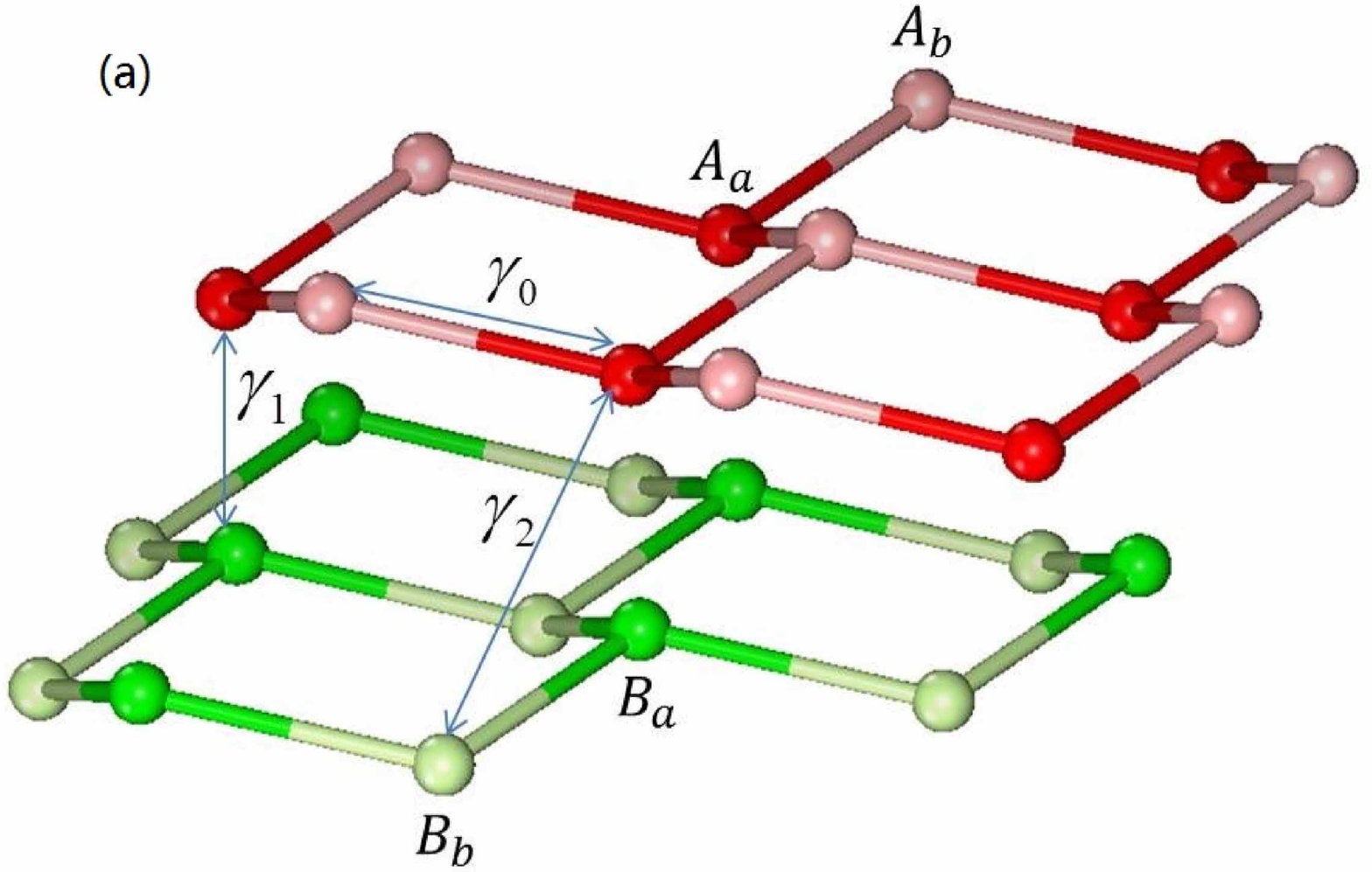}}}
\centerline{\scalebox{0.4}[0.4]{\includegraphics{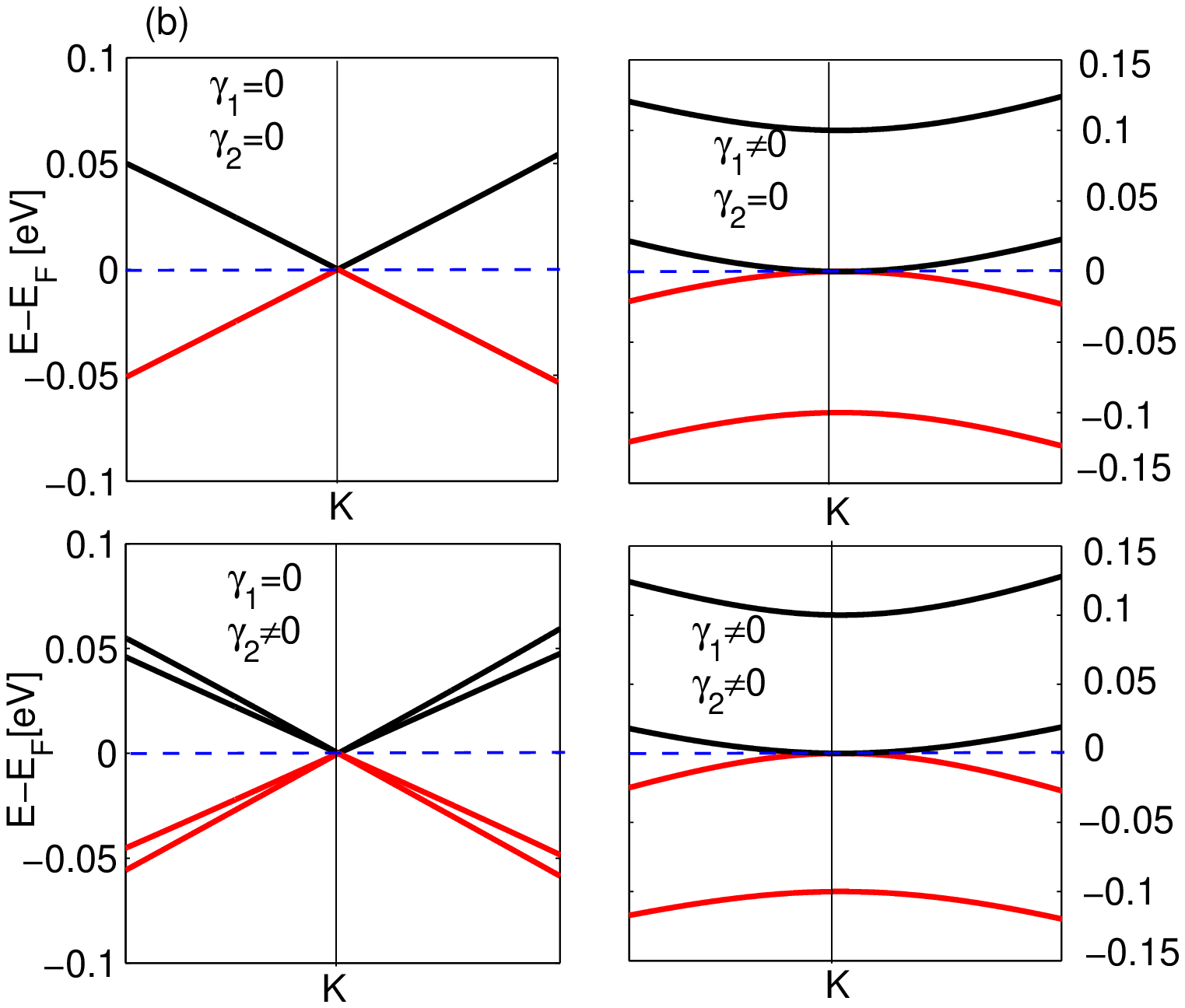}}}
\makeatother
\caption{\footnotesize{(Color online) (a) Lattice structure of bilayer silicene with AB stacking with the respective electronic hopping energies according to the TB Hamiltonian. (b) The energy-dispersion relations near the $K$ point with different of interlayer hopping energies $\gamma_{1}$ and $\gamma_{2}$ obtained from the TB approach.} }\label{fig4}

For $\gamma_{2}=0$, the vertical interlayer hopping $\gamma_{1}$ results in two parabolic bands with zero-gap, $E^{\pm}(\mathbf{q})=\pm{{v_{F}}^{2}q^2 }/{ \gamma_{1}}$, which touch at $E_{F}$ (as shown in FIGs. \ref{fig4}b), and mean that the bilayer silicene is metallic within this approximation. There are two additional bands that start at $\pm \gamma_{1}$, and the spectrum is electron-hole symmetric. If $\gamma_{1}=0$, the hopping $\gamma_{2}$ renormalizes the in-layer hopping $\gamma_{0}$, and there are four sets of Dirac-like linear bands while the velocity split into $v_{F}^{+}$ and $v_{F}^{-}$ with difference $\sqrt{3}a\gamma_{2}/{\hbar}$. However, the DFT calculations shows that the bands of bilayer silicene with AB stacking near the $K$ points maintain linear spectrums with the sole Fermi velocity. Hence, the interlayer hopping of the $\pi$ electrons in bilayer silicene doesn't influence the energy-dispersions near the $K$ points in the BZ and can be neglected. Comparing with the bilayer silicene, the bilayer graphene shows the parabolic bands with zero-gap near $K$ points due to the interlayer hopping intensely \cite{Latil,McCann,McCann1}. The planar configuration of in-layer C atoms can be maintained in bilayer graphene, so the $sp^2$ hybridization leads to the interlayer hopping easily. But for bilayer silicene, the buckled geometry promotes the mixing of $sp^{2}$ and $sp^3$ hybridization. The $sp^3$ hybridization results in the electrons tending locally. Especially, the buckling distance of bilayer silicene is larger that of monolayer silicene [see Table \ref{tab:structure}], and there are more components of $sp^3$ hybridization in bilayer silicene. Hence, the interlayer hopping of bilayer silicene can be ignored and the band structures feature Dirac-type electron dispersion in the vicinity of the corners of its hexagonal BZ.

In summary, we have demonstrated that bilayer silicene can remain stable in 2D honeycomb structure with AB stacking geometry from the first-principles calculations based on DFT.  The charge carriers in bilayer silicene behave like the massless Dirac fermions with linear energy dispersions near the $K$ points, similar to monolayer silicene. Within an insightful analysis of TB approach, we suggest that the buckled geometry, which is formed by mixing $sp^{2}$ and $sp^{3}$ hybridization, blocks the interlayer hopping and the Dirac-type electron dispersion is preserved in bilayer silicene.

This work was supported by the National Natural Science Foundation of China (11074313).


\newpage

List of figure caption

\begin{enumerate}

\item (Color online) Top view and Side view of bilayer silicene with (a) AA stacking and (b) AB stacking. The red pattern and green pattern represent the upper and lower layer of silicene, respectively. $\mathbf{a}_{1}$ and $\mathbf{a}_{2}$ denote the hexagonal lattice vectors, $d$ indicates the distance between two layers, and $t$ denotes the buckled distance. \label{fig1}

\item (Color online) Phonon-dispersion curves calculated by
supercell approach for 2D bilayer silicene (a)AA stacking and (b) AB stacking along the high-symmetry
line in the BZ. \label{fig2}

\item (Color online) The band structures of (a) monolayer silicene and (b) bilayer silicene with AB stacking show the remarkable linear dispersion of $\pi$ and $\pi^{*}$ bands at Fermi level. The dispersion-relations near the $K$ point are magnified at the right side. \label{fig3}

\item  (Color online) (a) Lattice structure of bilayer silicene with AB stacking with the respective electronic hopping energies according to the TB Hamiltonian. (b) The energy-dispersion relations near the $K$ point with different of interlayer hopping energies $\gamma_{1}$ and $\gamma_{2}$ obtained from the TB approach. \label{fig4}

\end{enumerate}

\end{document}